\documentclass[submission, Phys]{SciPost}

 \usepackage{bm,bbm,amsmath,mathrsfs}

 \newcommand* {\ket}[1]{\ensuremath{| {#1} \rangle}}
 \newcommand\redout{\bgroup\markoverwith{\textcolor{red}{\rule[.5ex]{2pt}{0.4pt}}}\ULon}
 
 \usepackage{hyperref}
 \hypersetup{colorlinks=true,citecolor=blue,linkcolor=red,urlcolor=blue}

 \definecolor{purple}{rgb}{0.5, 0., 0.9}

\usepackage{sidecap}
\graphicspath{{Figures/}}
\newcommand{\bl}{\begin{aligned}}
\newcommand{\el}{\end{aligned}}

\newcommand{\si}{\sigma}
\newcommand{\al}{\alpha}
\newcommand{\bt}{\beta}

\newcommand{\ga}{\gamma}
\newcommand{\la}{\lambda}

\newcommand{\vare}{\varepsilon}

\newcommand{\De}{\Delta}

\newcommand{\bsig}{{\bm\sigma}}
\newcommand{\tsig}{\tilde{\sigma}}

\newcommand{\ttau}{\tilde{\tau}}

\newcommand{\om}{i\omega_n}
\newcommand{\omp}{i\omega_{n'}}
\newcommand{\num}{i\nu_m}

\newcommand{\be}{\begin{equation}}
\newcommand{\ee}{\end{equation}}

\newcommand{\bea}{\begin{eqnarray}}
\newcommand{\eea}{\end{eqnarray}}
\newcommand{\bd}{\begin{displaymath}}
\newcommand{\ed}{\end{displaymath}}
\newcommand{\ba}{\begin{array}}
\newcommand{\ea}{\end{array}}
\newcommand{\bi}{\begin{itemize}}
\newcommand{\ei}{\end{itemize}}
\newcommand{\bc}{\begin{center}}
\newcommand{\ec}{\end{center}}
\newcommand{\bfl}{\begin{flushleft}}
\newcommand{\efl}{\end{flushleft}}
\newcommand{\bfr}{\begin{flushright}}
\newcommand{\efr}{\end{flushright}}

\newcommand{\hchi}{\hat{\chi}}

\newcommand{\hh}{\hat{h}}

\newcommand{\hG}{\hat{G}}
\newcommand{\hV}{\hat{V}}

\newcommand{\hbz}{\hat{{\bf z}}}

\newcommand{\hbk}{\hat{{\bf k}}}

\newcommand{\ua}{\uparrow}
\newcommand{\da}{\downarrow}

\newcommand{\tbq}{\tilde{{\bf q}}}
\newcommand{\tbQ}{\tilde{{\bf Q}}}

\newcommand{\tC}{\tilde{C}}

\newcommand{\fs}{\frac{1}{2}}

\def\ket#1{\left\vert #1 \right\rangle}
\def\dg{^{\dagger}}

\def\br{{\bf r}}
\def\bk{{\bf k}} \def\bK{{\bf K}}\def\bq{{\bf q}} 
  \def\bb{{\bf b}}
\def\bQ{{\bf Q}}\def\bg{{\bf g}} 
  
 \def\bd{{\bf d}}  
\def\bB{{\bf B}}  \def\hbz{\hat{{\bf z}}}

\def\da{\downarrow} \def\ua{\uparrow}
 
\def\dg{\dagger}

\def\bra{\langle}
\def\ket{\rangle}
\def\={\!\!\!&=&\!\!\!}
\def\+{\!\!\!&&\!\!\!+~}
\def\-{\!\!\!&&\!\!\!-~}

\begin{document}

\begin{center}{\Large \textbf{
 Magnetic excitations in the helical Rashba superconductor
 }}\end{center}

\begin{center}
 Alireza Akbari,  
Peter Thalmeier 
 \end{center}

 \begin{center}
Max Planck Institute for the  Chemical Physics of Solids, D-01187 Dresden, Germany
 \end{center}

 \begin{center}
 \today
 \end{center}

 \section*{Abstract}
 {\bf
We investigate the magnetic excitation spectrum in the helical state of a noncentrosymmetric superconductor  
with inversion symmetry breaking and strong Rashba spin-orbit coupling. For this purpose we derive the 
general expressions of the dynamical spin response functions under the presence of strong Rashba splitting of conduction bands,
superconducting gap  and external field which lead to stabilization of Cooper pairs with finite overall momentum
in a helical state. The latter is characterized by momentum space regions of paired and unpaired states with
different quasiparticle dispersions. The  magnetic response is determined by i) excitations within and between both paired
and unpaired regions ii) anomalous coherence factors and iii) additional spin matrix elements due to helical Rashba spin
texture of bands. We show that as a consequence typical correlated real space and spin space anisotropies appear in
the dynamical susceptibility which would be observable as a characteristic fingerprint for a helical superconducting state in inelastic 
neutron scattering investigations.
}

 \noindent\rule{\textwidth}{1pt}
 \tableofcontents\thispagestyle{fancy}
 \noindent\rule{\textwidth}{1pt}
 \vspace{10pt}

\section{Introduction}
\label{sec:introduction}

In this work we investigate the signatures of the helical state of noncentrosymmetric superconductors (NCS)
in the magnetic excitation spectrum obtained from inelastic neutron scattering (INS). 
NCS compounds  have broken inversion symmetry and consequently the gap functions in principle are mixtures of singlet and triplet components \cite{sigrist:09,takimoto:09}. This is also possible in 2D layered superconductors (SC) where inversion symmetry is broken in each layer although it is preserved overall in the 3D crystal \cite{goryo:12,akbari:14}.\\

In the normal state the inversion symmetry breaking leads to an odd Rashba spin orbit coupling that results in two nondegenerate split bands  $(\la=\pm1)$  with a momentum-locked spin texture characterized by opposite helicities. The corresponding Fermi wave numbers differ by an amount proportional to the size of Rashba spin-orbit coupling. The application of a field shifts the two Fermi spheres perpendicular to field direction and proportional to field strength.

In the superconducting state in an applied field this means that not only are singlet and triplet components mixed but also the Cooper pairs $(\bk+\bq~\si,-\bk+\bq~\si')$ will acquire a common pair momentum $2\bq$ proportional to the shift vector of Fermi surfaces and characterized by a gap function $\Delta^\bk_{\bq\la}\exp(i\bq\cdot\br)$. In this picture the orbital pair breaking is assumed to be small.  This commonly called `helical' state~\cite{kaur:05} is therefore of the Fulde-Ferrell (FF) \cite{fulde:64} type  but has a different composition of the condensation energy due to the effect of Rashba coupling than the original Zeeman-energy dominated FF case. The advantage of the helical state is that the finite pair momentum appears already at moderate fields due to the shifting of Rashba Fermi surface spheres. The modulus of its gap amplitude is, as in the FF state, constant in real space unlike the Larkin-Ovchinnikov (LO) state \cite{larkin:65} which has nodal planes. \\

Some aspects of the helical state including Rashba coupling and Zeeman term have been studied before, concerning mostly critical field curves~\cite{agterberg:07,loder:13,nakamura:15}. Microscopically
this state is characterized by a segmentation of Fermi surface sheets into paired and unpaired regions determined by the balance of kinetic, Rashba and Zeeman energies. The observation of this central aspect in the helical (and also in the original FF) state requires spectroscopic means. It has been proposed that STM-based quasiparticle interference (QPI) spectroscopy  may be used for FF\cite{akbari:16} and helical Rashba NCS \cite{akbari:22} cases. The complementary method of magnetic inelastic neutron spectroscopy (INS) has
also been suggested for  the FF state \cite{thalmeier:22}. It allows to investigate the signatures of finite momentum Cooper pairs  paired/unpaired segmentation of Fermi surface sheets on the magnetic excitation spectrum, in particular on the collective spin resonance formation within the superconducting gap and regarding the point group symmetry breaking anisotropy due to finite pair momentum.\\

In this work we generalize the analysis of magnetic excitations for the helical Rashba SC case. We first give a brief account of the periodic Rashba band model in the normal state. Then we discuss the simplest helical superconducting gap model introduced by Kaur et al \cite{kaur:05} and its condensation energy which determines the pair momentum $2\bq$ and gap size $\Delta^\bk_\bq$ as function of field strength.
In the main part of this work we derive the expressions for the dynamical magnetic susecptibility tensor which
contains the static response as well as the finite frequency magnetic spectrum. We investigate the field and momentum and polarization dependence of the latter which are in principle accessible by INS experiments.
Finally we discuss the temperature dependence of the static homogeneous susceptibility in view of the NMR Knight shift in the helical SC phase.

%
\begin{figure}[t]
\includegraphics[width=\linewidth]{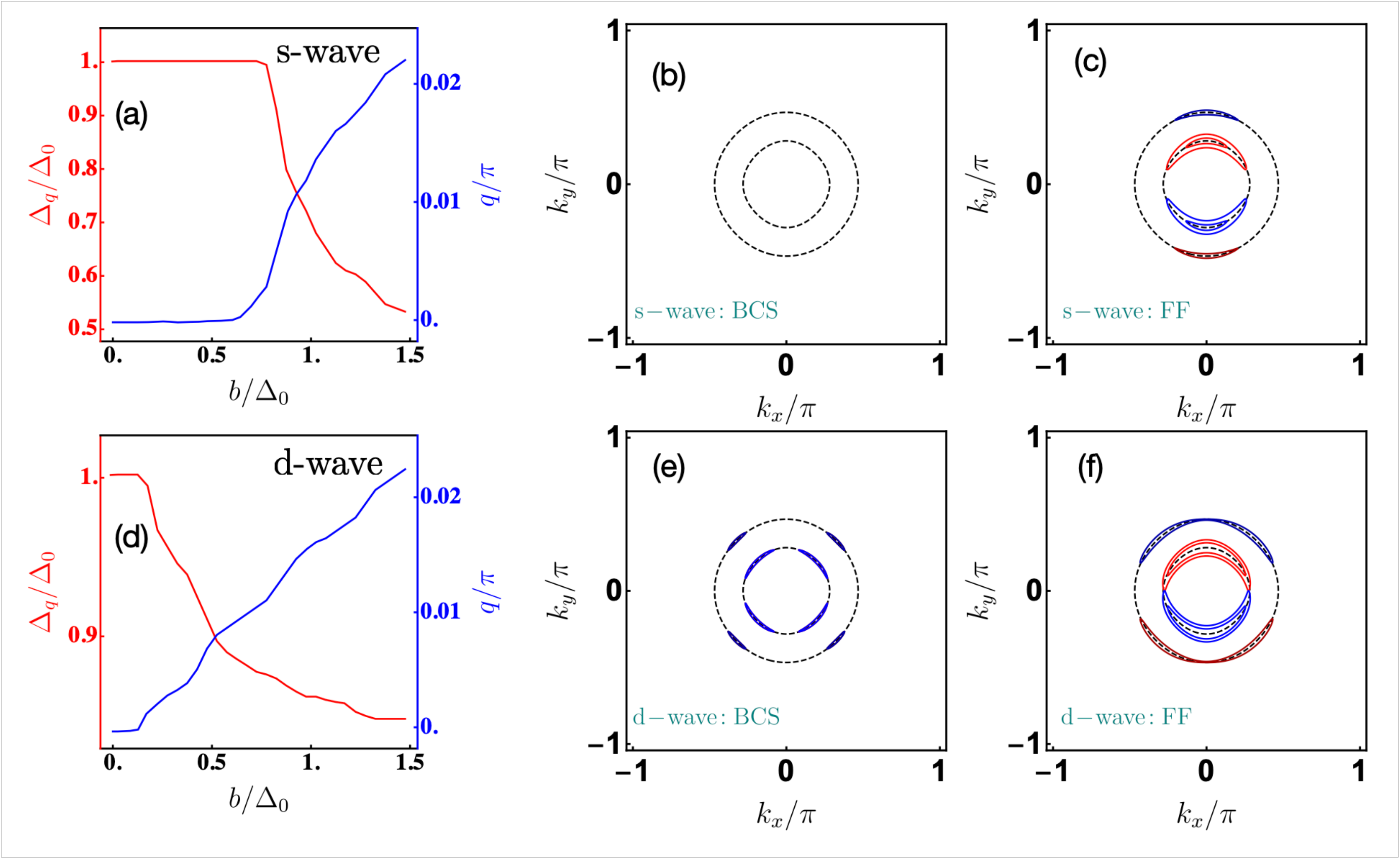}
\caption{Field dependence of pair momentum $q$ and gap amplitude $\Delta_q$ for small fields $(b\ll\alpha)$ (a,d).
Spectral functions i.e. quasipartilce sheets $|E_{\bk\bq\la}^\tau|=\omega$ in BCS case $(q=0)$ (b,e) for 
$\omega=0.06t<\Delta_0=0.2t$. 
In (b) quasiparticle sheets are absent while they appear around nodal directions in (e). In the helical case (c,f) $(b=0.9\Delta_0, q>0)$ extended sheets appear for both gap symmetries due to depairing centered around $k_y$ direction perpendicular to $\bb=b\hat{{\bf x}}$. The blue/red sheets correspond to $\tau=+/-$ and outer/inner sheets to $\lambda=+/-$. In all figures the designations  ``s,d- wave" refer to the 
gap form factors $f_\Gamma(\bk)$ of the inversion symmetric limit where $\alpha =0$ (Sec.~\ref{subsect:supermodel}).}
\label{fig:spectral}
\end{figure}
%

\section{Normal state Rashba bands and states}
\label{sec:model}
Here we introduce the widely used bandstructure model  including the Rashba coupling originating from inversion-symmetry breaking \cite{sigrist:09,akbari:22}. We use the periodic form in view of the later calculations of magnetic response functions but occasionally discuss the features of Rashba bands in the  the convenient parabolic approximation. 
 In the spin representation, the 2D Rashba Hamiltonian in an external field is characterized by the following~\cite{kaur:05}
\be
\bl
&
H_{0}=
\sum_\bk\Psi^\dag_\bk h_{0\bk}\Psi_\bk; \;\;\;
 h_{0\bk}=\xi_\bk\sigma_0+(\al\bg_\bk+\bb)\cdot\bsig 
 .
\label{eq:HRashba}
\el
\ee
Here $\Psi^\dag_\bk=(a^\dag_{\bk\ua},a^\dag_{\bk\da})$ represent conduction electrons with the tight binding (TB) dispersion $\vare_\bk=-2t(\cos k_x+\cos k_y)$ with $-\pi\leq k_x,k_y\leq \pi$. Furthermore $t>0$ is the hopping element corresponding to a conduction band half-width $D_c=4t$ and $\xi_\bk\equiv\vare_\bk-\mu_{\rm TB}$ . The chemical potential $\mu_{\rm TB}$ falls in the interval $-D_c\leq \mu_{\rm TB}\leq D_c$ and  is counted from the {\it band center} $\vare_\bk=0$. It is useful to connect this to the 2D parabolic band model for $\mu_{\rm TB}\leq 0$ with $\vare_\bk=\vare_0+\bk^2/2m$. With $\vare_0=-D_c$ denoting the band bottom and $m=2/D_c$ the effective mass. The chemical potential counted from the {\it band bottom} is then obtained by $\mu_P=\mu_{\rm TB}-\vare_0 \geq 0$. Furthermore  $\bb=\mu_B{\bf B}$ is the energy scale of the applied magnetic field \bB.\\
   
The Rashba spin-orbit coupling is odd under inversion with  $\bg_{-\bk}= -\bg_{\bk}$, explicitly $\bg^P_\bk=(k_y,-k_x,0)/k_F = (\sin\theta_\bk,-\cos\theta_\bk,0)$ in the parabolic model. Here $\theta_\bk$ is the azimuthal angle of $\bk$ measured in relation to the $k_x$- axis. Moreover,  $k_F=(2m\mu)^\fs$ is the Fermi wave number and $v_F=k_F/m$ is the corresponding velocity. To retain consistency with the TB dispersion we will employ the periodic form $\bg^{\rm TB}_\bk=(\sin k_y,-\sin k_x,0)$. Both forms are normalized according to  $|\bg^P_\bk|=1$ and  $|\bg^{\rm TB}_\bk|_{\rm max}=\sqrt{2}$. Equivalency for $k_x,k_y\ll\pi$ requests that $\alpha_P=k_F\alpha_{\rm TB}$. We will discard the indices TB, P from now on and mostly rely on the context. The diagonalised Hamiltonian of Eq.~(\ref{eq:HRashba}) reads %
\be
\bl
&
H_{0}
=\sum_{\bk\la}\vare_{\bk\la} c^\dg_{\bk\la}c_{\bk\la};\;\;\;
\vare_{\bk\la}(\bb)
=\xi_{\bk}+\la|\al\bg_{\bk}+\bb|,
\label{eq:Rdispb1}
\el
\ee
Here $\vare_{\bk\la}(\bb)$ are Rashba- split and Zeeman- shifted bands (energies counted from $\mu$) with band states corresponding to helicities $\la=\pm 1$. For vanishing field the two Rashba bands may be written as
\be
\vare^0_{\bk\la}=\xi_{\bk}+\la |\al\bg_{\bk}| = \frac{1}{2m}(k+\la k_0)^2-\tilde{\mu},
\label{eq:Rdisp0}
\ee
with 
$k_0=\fs\frac{|\alpha|}{\mu}k_F; \; {\rm and} \; \tilde{\mu}=\mu(1+\frac{1}{4}\frac{\al^2}{\mu^2})$. 
These two parabolic dispersions are shifted by an amount $k_0$. The two resulting Fermi spheres have approximate radii $k_F^\la=k_F-\la k_0=k_F(1-\frac{\la}{2}\frac{|\alpha|}{\mu})$ for  moderately small Rashba coupling strength $|\alpha|\ll\mu$. Then their relative deviation $(k_F^--k_F^+)/k_F=|\alpha|/\mu$ is a direct measure for the size of $\alpha$. We assumed a physically appropriate hierarchy of energy scales described by  $(b < |\alpha| < \mu < D_c)$.\\

%
\begin{figure}[t]
\includegraphics[width=\linewidth]{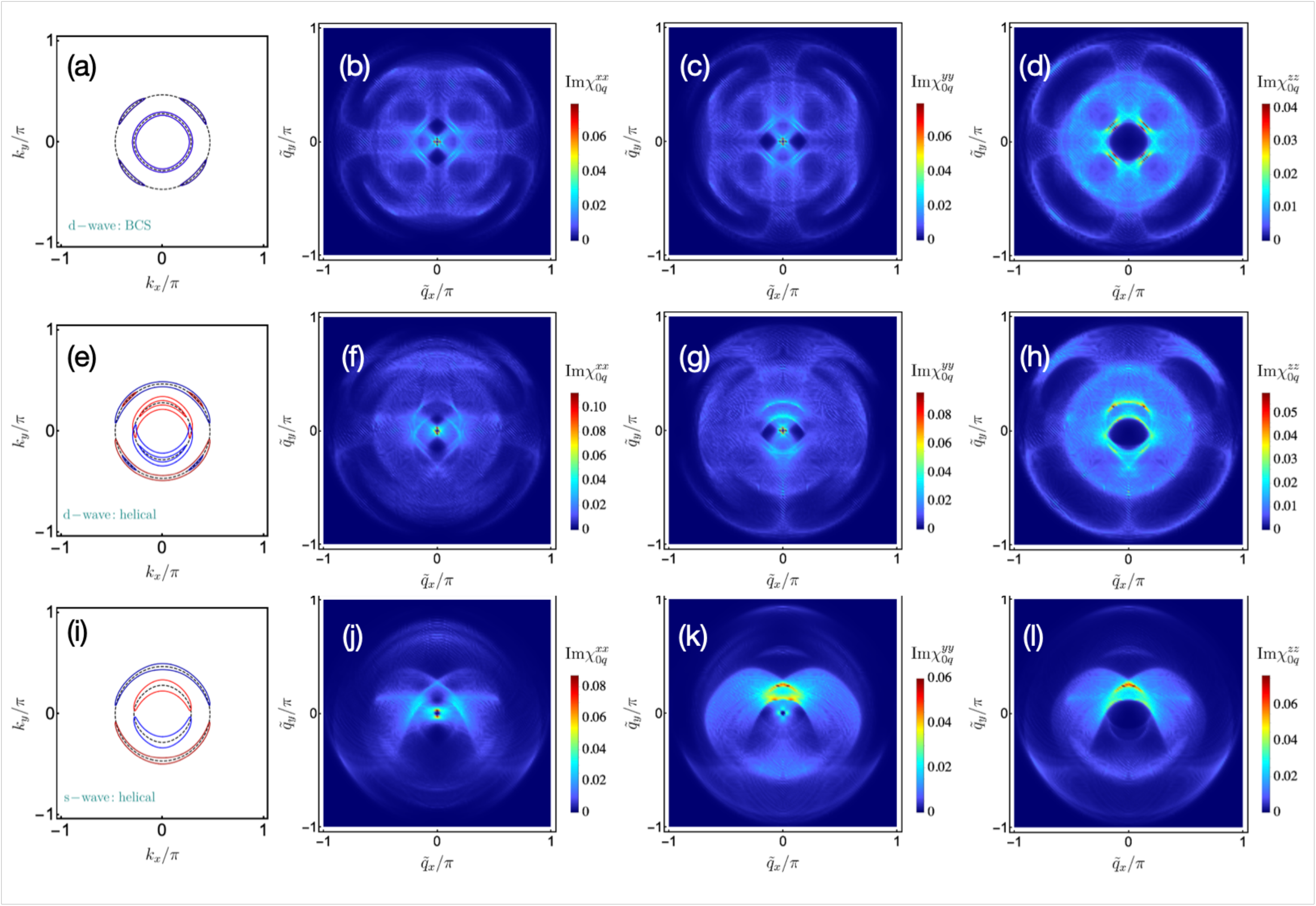}
\caption{Brillouin-zone cuts of spectral functions (a,e,i) and corresponding dynamical magnetic structure function for 
$x,y,z$ polarisation in the BCS d-wave (b-d), helical d-wave (f-h) and helical s-wave (j-l) cases for 
$\omega=0.12t<\Delta_0=0.2t$.
In the BCS cases (b-d) the $ xx$ and $yy$ response is related by a rotation $R_\frac{\pi}{2}$ around the $z$ axis while
the $zz$ response is by itself fourfold symmetric (Eq.~(\ref{eq:BZsymmetry})). These symmetries are lost in the hellical phase for both gap models
due to the distinguished common pair momentum oriented along $q_y$.}
\label{fig:BZcut}
\end{figure}
%

For finite but small field $\bb$ parallel to the plane
the Rashba Fermi surfaces resulting from Eq.~(\ref{eq:Rdispb1})
are shifted {\it perpendicular} to the field orientation in opposite directions \cite{sigrist:09,akbari:22} by a shift vector
\bea
q_s=
\frac{b}{2\mu}k_F=\frac{m\mu_BB}{k_F}=\frac{b}{v_F}.
\label{eq:Rshift}
\eea
In brief, the splitting of Rashba sheets is a direct measure for the Rashba coupling strength $|\alpha|$ while their shifting perpendicular to \bB~is a determined only by field strength. We will use $\bb=b\hat{{\bf x}}$ and $\bq_s=q_s\hat{{\bf y}}$ for the geometric configuration.
Finally we give the unitary transformation from spin states to the helical Rashba eigenstates  $|\bk\la\rangle=c^\dag_{\bk\la}|0\rangle$  which are created by  operator pairs $\Phi_\bk^\dag=(c_{\bk +}^\dag c^\dag_{\bk -})$ $(\la=\pm)$. It is defined by \cite{thalmeier:20} 
\bea
\Phi^\dag_\bk=\Psi^\dag_\bk S_\bk; \;\;\;
S_\bk=
\frac{1}{\sqrt{2}}
\left[
\begin{matrix}
1& ie^{-i\theta_\bk} \cr
 ie^{i\theta_\bk} &  1
 \end{matrix}
 \right],
 \label{eq:helical}
\eea
where the angle is obtained from 
$$\theta_\bk=-\tan^{-1}(g_{\bk x}/g_{\bk y})=\tan^{-1}(\sin k_y/\sin k_x)\rightarrow \tan^{-1}(k_y/k_x).$$
 Here the second and last expressions correspond to TB and parabolic bands, respectively. In the latter $\theta_\bk$ is simply the azimuthal angle of \bk. \\

\section{Helical superconductor: Segmentation into paired and unpaired regions}
\label{subsect:supermodel}

In the present work we do not investigate possible mechanisms of superconductivity in NCS  compounds, for an extended review see Ref.~\cite{sigrist:09}. In the compounds with Rashba spin-orbit coupling phonons~\cite{wiendlocha:16} and alternatively spin-fluctuations~\cite{yanase:08,takimoto:09,mukherjee:12} may provide a mechanism for Cooper pair formation. 

Here we investigate the possibility of a superconducting state with common  momentum $2\bq$ of Cooper pairs due
to the pairbreaking effect of the external field in combination with Rashba spin-orbit coupling and how this will influence
the dynamical magnetic response. The actual $\bq=q\hat{{\bf y}}$ has to be evaluated by the minimization of the condensation energy in the helical SC phase given below, one may expect it to be correlated with the Rashba shift vector $\bq_s=q_s\hat{{\bf y}}$ in Eq.~(\ref{eq:Rshift})\cite{yanase:07,nakamura:15}.
More general inhomogeneous SC phases with several $\bq_i$, e.g. the 
`stripe phase'~\cite{kaur:05} with the pair $(\bq,-\bq)$ will not be discussed here. Furthermore we restrict the choice of the helical SC gap functions $\Delta_{\bq\la}(\bk)$ to the minimal model proposed by Kaur et al~\cite{kaur:05} whose essentials we briefly mention here: It assumes that in the limit of $\al\rightarrow 0$ the SC gap is of the {\it singlet} type (e.g. s-,d-wave) characterized by an orbital basis function $f_\Gamma(\bk)$ belonging  to the irreducible representation of the tetragonal symmetry group $D_{4h}$. Turning on a finite $\alpha$ leads to an additional \bk-dependence of effective interaction and gap function due to the admixture of triplet components enforced by the helical spin structure. For small fields $b\ll|\al |$ this \bk-dependence may be eliminated by a phase transformation $\Delta_{\bq\pm}^{\Gamma\bk}\rightarrow\pm\exp(\mp i\theta_\bk)\Delta_{\bq\pm}^{\Gamma\bk}$ (for simplicity we keep the same symbol for the transformed gap function). It is associated with a correspondingly transformed effective interaction in helicity representation given by\cite{kaur:05}
\bea
\hV=-  \frac{V_\Gamma(\bk\bk')}{2}(\sigma_0-\sigma_x); \quad V_\Gamma(\bk\bk')=V_0^\Gamma f_\Gamma(\bk)f_\Gamma(\bk').
\label{eq:vmat0}
\eea
Inserting this two-band pairing interaction into the gap equation leads to the condition $\Delta^\bk_{\bq-}=-
\Delta^\bk_{\bq+}$ \cite{kaur:05,akbari:22}. The opposite sign of the two gaps is enforced by the opposite spin texture on the two Rashba bands. For the later numerical discussion we will consider the isotropic s-wave case $f_\Gamma(\bk)=1$ and the d-wave case  $f_\Gamma(\bk)=(\cos k_x-\cos k_y)$ correspoinding to  $\Delta^\bk_{\bq\la}=\Delta_{\bq\la}f_\Gamma(\bk)$.\\

Introducing the Nambu spinors $\psi^\dg_{\bk\bq\la}=(c^\dg_{\bk+\bq\la},c_{-\bk+\bq\la})$ 
the total BCS Hamiltonian is given by%
\be
\bl
\hat{H}_{\rm BCS}
=\frac{1}{2}\sum_{\bk\la}\psi^\dg_{\bk\bq\la}\hh_{\bk\bq\la}\psi_{\bk\bq\la} + 
\frac{1}{2}\sum_{\bk\la}\vare_{\bk+\bq\la} +
\frac{1}{2}\sum_{\bk\la}\frac{\Delta^{\bk 2}_{\bq\la}}{V_0}
\el
\ee
with  the Hamilton matrix represented by
\be
\bl
\hh_{\bk\bq\la}=
\vare^a_{\bk\bq\la}\tau_0+
\left[
 \begin{array}{cc}
 \vare^s_{\bk\bq\la}&-\Delta^\bk_{\bq\la} \\
 -\Delta_{\bq\la}^{\bk *}& -\vare^s_{\bk\bq\la}
\end{array}
\right]
.
\label{eq:hsmat}
\el
\ee
Considering the symmetries $\xi_\bk=\xi_{-\bk}$ and $\bg_\bk=-\bg_{-\bk}$ the diagonal matrix elements may be written as
\be
\bl
\vare_{\bk +\bq\la}(\bb)
&=
\xi_{\bk +\bq}+\la|\al\bg_{\bk+\bq}+\bb|
,
\\
\vare_{-\bk +\bq\la}(\bb)
&=
\xi_{\bk -\bq}+\la|\al\bg_{\bk-\bq}-\bb|
.
\el
\ee
It is convenient to introduce symmetric (s) and antisymmetric (a) expressions according to
%
%
%
\be
\bl
\vare^{s,a}_{\bk\bq\la}
&=
\frac{1}{2}(\vare_{\bk +\bq\la}\pm\vare_{-\bk +\bq\la})
.
\label{eq:symeps}
\el
\ee
They have even/odd symmetry $\vare^{s}_{-\bk\bq\la}=\vare^{s}_{\bk\bq\la}$ and
$\vare^{a}_{-\bk\bq\la}=-\vare^{a}_{\bk\bq\la}$ with respect to inversion. The latter enforces
the property $\sum_{\bk\la}\vare^a_{\bk\bq\la}=0$ where the sum over \bk~covers both paired and unpaired regions explained below.\\
The second term  in Eq.~(\ref{eq:hsmat})  can be diagonalized by a Bogoliubov transformation \cite{cui:06,akbari:22} leading to quasiparticle states created
by $\alpha_{\bk\la},\beta_{\bk\la}$ and a corresponding quasiparticle Hamiltionian
\bea
\bl
H_{\rm BCS}
= &
E_G+ \frac{1}{2}\sum_{\bk\la}(|E^+_{\bk\bq\la}|\al^\dg_\bk\al_\bk+|E^-_{\bk\bq\la}|\bt^\dg_\bk\bt_\bk)
.
\label{eq:bcsbogol}
\el
\eea
To visualize the quasiparticle sheets we will use the spectral function 
\bea
\bar{A}^\la_{\bk\bq}(\omega >0)=\delta(\omega-|E^+_{\bk\bq\la}|)+\delta(\omega-|E^-_{\bk\bq\la}|)
\eea
in subsequent figures. Here the (positive) quasiparticle energies $|E^\tau_{\bk\bq\la}|$ are given as $(\tau=\pm,\bar{\tau}=\mp)$: 
\bea
\bl
E^\tau_{\bk\bq\la}
&=E_{\bk\bq\la}+\tau\vare^a_{\bk\bq\la}=E^{\bar{\tau}}_{-\bk\bq\la},
\\
E_{\bk\bq\la}
&=[\vare^{s2}_{\bk\bq\la}+\Delta_{\bq\la}^{\bk 2}]^\frac{1}{2}=E_{-\bk\bq\la}.
\label{eq:SCquasi}
\el
\eea
If both $E^\tau_{\bk\bq\la}>0$  $(\tau=\pm)$ for a specific \bk~ and $\la$ the Cooper pair state with pair momentum $2\bq$ is stable. However, if $E^+_{\bk\bq\la}<0$ or $E^-_{\bk\bq\la}<0$ the pair state is instable and only unpaired  quasiparticle states exist for the wave vectors $\bk+\bq$, $-\bk+\bq$. Although for such wave vectors $|E^\pm_{\bk\bq\la}|$ are normal quasiparticle excitations their energy nevertheless depends on the gap size $\Delta_{\bq\la}$ determined only by the paired FS sections. This is due to the fact that in the coherent helical SC ground state the unpaired electrons and holes also experience the pairing potential supported by the paired electrons, although they don't contribute to it.\\

%
%

The constant $E_G=\langle H_{\rm BCS}\rangle$ appearing in Eq.~(\ref{eq:bcsbogol})  is equal the total ground state  energy $E_G(\bq,\Delta_{\bq\pm})$ of the helical  state. Subtracting the  normal state ground state energy $E^0_G=(1/2)\sum_{\bk\la}(\vare^0_{\bk\la}-|\vare^0_{\bk\la}|)$
we obtain the superconducting condensation energy $E_c=E_G-E^0_G$ as \cite{akbari:22}
\bea
\bl
E_c(\bq,\Delta_{\bq\pm})=
\fs\sum_\la
\Bigg[
&
\sum_\bk
N\bigl(\frac{|\Delta^\bk_{\bq\la}|^2}{V_0}\bigr)
-\sum_\bk
\Bigl\{(E_{\bk\bq\la}-|\vare^0_{\bk\la}|)
+(\vare^s_{\bk\bq\la}
\!-\!
\vare^0_{\bk\la})
\\
&
\!
+[E^+_{\bk\bq\la}\Theta(-E^+_{\bk\bq\la})+E^-_{\bk\bq\la}\Theta(-E^-_{\bk\bq\la})]
\Bigr\}
\Bigg],
\el
\label{eq:condens}
\eea
where
$\Delta^{\bk}_{\bq\la} =\Delta_{\bq\la}f_{\Gamma}(\bk)$.
 In both s,d-wave cases the normalization  is
 $(1/N)\sum_{\bk} f_{\Gamma}(\bk)^2=1$
 in the first $\bk$-sum of the above equation so that this term is equal to $N|\Delta_{\bq\la}|^2/V_0$.
Because of the separation of Eq.~({\ref{eq:symeps}) the odd $\vare^a_{\bk\bq\la}$ Rashba energies enter only in the last term of Eq.~(\ref{eq:SCquasi}) but not under the square root. The above energy functional  must be minimized with respect to \bq~and $\Delta_{\bq\pm}$ for a given size of the Rashba coupling $\alpha$ and as function of field $b$.
Possible ground states are the helical SC state  $(\bq\neq 0,|\Delta_{\bq\la}|>0)$, the BCS state $(\bq=0,|\Delta_{0\la}|>0)$ or the unpolarized normal $(b=0,\bq=0, \Delta_{\bq\la}=0)$ states. We note again that we are restricted to the small field range $b\ll|\alpha|$ due to the 
assumption of a field-independent spin texture only determined by the Rashba term.\\ 

The minimization problem is much simplified  by the equal gap magnitude $|\Delta_{\bq\pm}|=\Delta_\bq$ in the model of Eq.~(\ref{eq:vmat0}). Strictly speaking this holds only when \bq=0 but this minimization constraint will also be kept for the helical SC case. The effective interaction strength $V_0$ in Eqs.~(\ref{eq:vmat0},\ref{eq:condens}) is connected to the BCS gap amplitude $\Delta_0$ by the gap equation $(b=0)$:
\bea
\frac{1}{V_0}=\frac{1}{2N}\sum_{\bk\la}\frac{f_\Gamma(\bk)^2}{2E_{\bk\la}}.
\label{eq:pairpot}
\eea
Here the BCS zero-field quasiparticle energy is simply $E_{\bk\la}=[\vare^{02}_{\bk\la}+\Delta_0^2]^\fs $. 
Minimization of $E_c(\bq,\Delta_{\bq})$  with respect to $\Delta_{\bq}$ and \bq~ determines the true gap $\Delta_\bq(b,\al)$ and wave vector  $\bq(b,\al)$ that characterise the helical SC state. One must keep in mind, however, that the model defined in Eq.~(\ref{eq:vmat0}) is only valid in the low field limit  $b/|\alpha | <1$.  An example of the resulting $\Delta_q(b), q(b)$ dependence for small fields and fixed $\alpha$ is shown in Fig.~\ref{fig:BZcut}(a,d) for the two gap symmetries. These curves depend considerably on the chemical potential $\mu_P$ and Rashba coupling $\alpha$ because for numerical reasons the gap value $\Delta_0$ is not negligible compared to them.

%
\begin{figure}
\includegraphics[width=\linewidth]{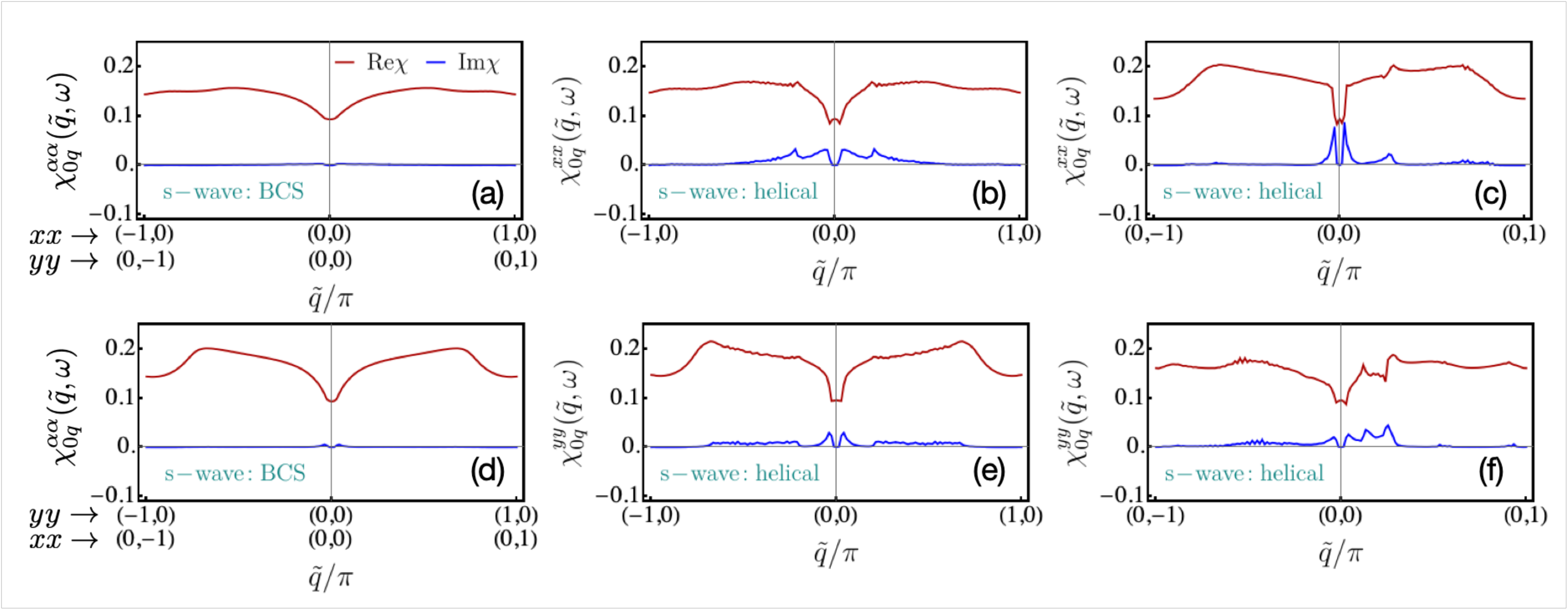}
\caption{Axis cuts along $(10)$ and $(01)$ of $xx, yy$ dynamical magnetic structure functions for s-wave case at $\omega=0.12t$.
In the BCS case $(q=0)$ the $xx(10)$ and $yy(01)$ pairs (a) and likewise $yy(10)$ and $xx(01)$ pairs (d) are equivalent due to $R_\frac{\pi}{2}$ rotational symmetry. In the helical phase this symmetry is lost due to distinguished $q_y$ direction of common pair momentum which introduces combined momentum space and spin space anisotropy. This is obvious from comparison of (a) with (b,c) and (d) with (e,f)    
(see also Fig.~\ref{fig:BZcut}).}
\label{fig:linecut}
\end{figure}
%

\section{The dynamical magnetic response function for the helical Rashba superconducting state}
\label{sec:helical-response}

Now we come to the main objective of this work. The calculation of the dynamical magnetic response function of a helical Rashba superconductor is considerably more involved than in the  simple BCS state \cite{bulut:96,norman:00} or even the centrosymmetric FF superconductor \cite{mier:09,thalmeier:22} due to the complicated spin textures of Rashba bands which is encoded in the 
unitary transformation matrix to helical states given in Eq.(\ref{eq:helical}). Here we give the details of its derivation.
For the noninteracting helical Rashba quasiparticles the dynamical magnetic susceptibility is obtained from the bubble diagram without vertex corrrections. Because of the helical spin structure in principle all cartesian elements $(\alpha,\beta=x,y,z)$ of the susceptibilities may be nonzero and different. They are defined by
\bea
\chi_{\al\bt}(\tbq,\num)=\frac{1}{N}\int_0^{\tilde{\bt}}d\ttau e^{\num\ttau}\bra T_{\ttau} S_\al(\tbq,\ttau)S_\bt^\dag(\tbq,0)\ket
,
\eea
where $\tilde{\bt}=1/kT$ and the spin operators are given in spin $(a_{\bk\si}$) and $(c_{\bk\la}$) helical bases as
\bea
S_\alpha(\tbq)=\fs\sum_{\bk\si\si'}a^\dag_{\bk+\tbq\si'}\si^\al_{\si'\si}a_{\bk\si}
=\fs\sum_{\bk\la\la'}c^\dag_{\bk+\tbq\la'}\tsig^\al_{\la'\la}(\bk+\tbq,\bk)c_{\bk\la}
.
\eea
Here the latter presentation has to be chosen as it forms the eigenbasis of the Rashba Hamiltonian. Therefore the effective spin operators $\fs\tsig_{\la'\la}^\alpha(\bk+\tbq,\bk)$ in this basis are now momentum dependent and they are obtained by the transformation of Eq.(\ref{eq:helical}) according to
\bea
\tsig_{\la'\la}^\alpha(\bk',\bk)=\sum_{\si,\si'}S^*_{\si'\la'}(\bk')\si^\al_{\si'\si}S_{\si\la}(\bk).
\label{eq:effspin}
\eea
The matrices (with $\la'\la$ indices) have the conjugation property $\tsig^\al(\bk',\bk)^\dg =\tsig^\al(\bk,\bk')$ and obey the commutation rules
$[\tsig^{\al}(\bk',\bk),{\tsig^{\bt}(\bk',\bk)}^\dg]=i\epsilon_{\al\bt\ga}(\tsig^\ga(\bk')+\tsig^\ga(\bk))$ where $\epsilon_{\al\bt\ga}$ is the fully antisymmetric tensor and $\tsig^\ga(\bk)\equiv\tsig^\ga(\bk,\bk)$ etc. The cartesian spin expectation values in each Rashba state $|\bk\la\ket$ are given by $\bra\tilde{\bsig}^\la(\bk)\ket = (-\la\sin\theta_\bk,\la\cos\theta_\bk,0)$. Therefore the spins of Rashba states are perpedicular to the momentum direction $\hbk=(\cos\theta_\bk,\sin\theta_\bk,0)$, i.e. $\bra\tilde{\bsig}^\la(\bk)\ket\cdot\hbk=0$ and $\bra\tilde{\bsig}^\la(\bk)\ket\times\hbk=\la\hbz$. Furthermore they are opposite on the two Rashba bands $\la=\pm$.\\

Using the helical eigenbase and its corresponding quasiparticle Green's functions and effective spin operators the dynamical susceptibility may now be written as
\bea
\chi^{\al\bt}_{0\bq}(\tbq,\num) 
=-\frac{T}{4}\frac{1}{N}\sum_{\bk n\la\la'}\tsig^\al_{\la\la'}(\bk',\bk)\tsig^\bt_{\la'\la}(\bk,\bk')\sum_nTr_\tau[\hG_\bq(\bk,\om)\hG_\bq(\bk',\om')],
\label{eq:dynsus1}
\eea
where $\omp=\om+\num$ with $n'=n+m$. 
Here $\tau$ is the Nambu index in particle-hole space and $Tr_\tau$ denotes the corresponding trace of the product of Nambu Green's function matrices given in Eq.~(\ref{eq:gmat}).
The latter are  obtained from Eq.~(\ref{eq:hsmat}) as
\be
\bl
\hG_{\bq\la}(\bk,\om) = \;&
(\om-\hh_{\bk\bq\la})^{-1}
\\
 =\;
&
 \frac{1}{(\om-E^+_{\bq\bk\la})(\om+E^-_{\bq\bk\la})}
 \left[
 \begin{array}{cc}
\om+\vare^s_{\bk\bq\la}- \vare^a_{\bk\bq\la}
&-\Delta^\bk_{\bq\la} \\[0.2cm]
-\Delta^{\bk *}_{\bq\la}
 &\om -\vare^s_{\bk\bq\la}- \vare^a_{\bk\bq\la}
\end{array}
\right].
\label{eq:gmat}
\el
\ee
Using the definitions
\bea
\bl
&
M^{\al\bt}_{\la'\la}(\bk',\bk)=\tsig^\al_{\la\la'}(\bk',\bk)\tsig^\bt_{\la'\la}(\bk,\bk');\;\;\;
\\
&
\hchi_{\la\la'}(\bk\bk',\num)=-\frac{T}{4}\sum_nTr_\tau[\hG_\bq(\bk,\om)\hG_\bq(\bk',\omp)],
\label{eq:effmat}
\el
\eea
the response function may be written as a product of helical state spin matrix elements and a dynamical kernel, respectively, according to
$(\num\rightarrow \omega+i\eta)$ 
\bea
\chi^{\al\bt}_{0\bq}(\tbq,\omega) 
=\frac{1}{N}\sum_{\la\la'}
\sum_{\bk}M^{\al\bt}_{\la'\la}(\bk',\bk)\hchi_{\la\la'}(\bk\bk',\omega)
.
\label{eq:dynsus2}
\eea
First we calculate the kernel where the trace may be evaluated as
\be
\fs Tr_\tau[\hG_\bq(\bk,\om)\hG_\bq(\bk',\omp)]=\frac
{(\om-\vare^a_{\bk\bq\la})(\omp-\vare^a_{\bk'\bq\la'})+\vare^s_{\bk\bq\la}\vare^s_{\bk'\bq\la'}+\Delta_{\bq\la}^\bk\Delta_{\bq\la'}^\bk}
{(\om-E^+_{\bk\bq})(\om+E^-_{\bk\bq})(\omp-E^+_{\bk'\bq})(\omp+E^+_{\bk'\bq})}
.
\ee

Performing the sum over the Matsubara frequencies $\omega_n$ and analytically continuing to the real axis
according to $\num\rightarrow \omega+i\eta$ a lengthy calculation leads to the
final result for the kernel in the dynamical susceptibility of Eq.~(\ref{eq:dynsus2})
\be
\bl
\hchi_{\la\la'}(\bk\bk',\omega) =
&
\fs\tC^\bq_+(\bk\la,\bk'\la')\bigl[
\frac{f(E^+_{\bk'\bq\la'})-f(E^+_{\bk\bq\la})}{\omega-(E^+_{\bk'\bq\la'}-E^+_{\bk\bq\la})+i\eta}-
\frac{f(E^-_{\bk'\bq\la'})-f(E^-_{\bk\bq\la})}{\omega+(E^-_{\bk'\bq\la'}-E^-_{\bk\bq\la})+i\eta}
\bigr]+
\\
&
\fs\tC^\bq_-(\bk\la\bk'\la')\bigl[
\frac{1-f(E^-_{\bk'\bq\la'})-f(E^+_{\bk\bq\la})}{\omega+(E^-_{\bk'\bq\la'}+E^+_{\bk\bq\la})+i\eta}+
\frac{f(E^+_{\bk'\bq\la'})+f(E^-_{\bk\bq\la})-1}{\omega-(E^+_{\bk'\bq\la'}+E^-_{\bk\bq\la})+i\eta}
\bigr]
,
\label{eq:kernel-sus}
\el
\ee
where $f(E)=(\exp(E/T)+1)^{-1}$ is the Fermi function. We may also obtain a different presentation of the last two terms by using $1-f(E)=f(-E)$.
The anomalous superconducting coherence factors of magnetic response for
the helical phase are given by
\bea
\bl
\tC^\bq_\pm(\bk\la\bk'\la')=\fs\bigl[
1\pm \frac{\vare^s_{\bk\bq\la}\vare^s_{\bk'\bq\la'}+\Delta_{\bq\la}^\bk\Delta_{\bq\la'}^{\bk'}}
{E_{\bk\bq\la}E_{\bk'\bq\la'}}\bigr]
.
\label{eq:cohfac}
\el
\eea
Furthermore we have to compute the matrix elements in Eq.~(\ref{eq:effmat}) for the evaluation of the
 susceptibility of Eq.~(\ref{eq:dynsus2}). It is useful to note that they satisfy Hermitean symmetry  which derive from the fact that the effective spin operators in helical representation are also Hermitean, i.e. fulfil
 $\tsig^\al_{\la'\la}(\bk',\bk)=\tsig^{\al *}_{\la\la'}(\bk\bk')$. Then it follows that
 \bea
 \bl
 M^{\al\bt}_{\la'\la}(\bk,\bk')
 =
 \;
 & 
 M^{\bt\al}_{\la\la'}(\bk\bk')^*
 ,
 \\
 M^{\al\al}_{\la'\la}(\bk,\bk')
 =
 \;&
 M^{\al\al}_{\la\la'}(\bk\bk')^*=|\tsig^\al_{\la\la'}(\bk\bk')|^2
 ,
 \label{eq:matsymm}
 \el
 \eea
so that cartesian diagonal elements are real symmetric. Using the explicit form given by Eqs.~(\ref{eq:effspin},\ref{eq:helical}) we derive the latter $(\la'\la)$ matrices as
\bea
\bl
\left\{
\begin{array}{r}
M^{xx}(\bk'\bk)  \\
M^{yy}(\bk'\bk)
\end{array}
\right\}
&=
\fs\left(
\begin{matrix}
1\mp\cos(\theta_\bk+\theta_{\bk'})& 1\pm\cos(\theta_\bk+\theta_{\bk'}) \cr
 1\pm\cos(\theta_\bk+\theta_{\bk'}) &  1\mp\cos(\theta_\bk+\theta_{\bk'})
 \end{matrix}
 \right),
 \\
 M^{zz}(\bk'\bk)
 &=
\fs \left(
\begin{matrix}
1-\cos(\theta_\bk-\theta_{\bk'})& 1+\cos(\theta_\bk-\theta_{\bk'}) \cr
 1+\cos(\theta_\bk-\theta_{\bk'}) &  1-\cos(\theta_\bk-\theta_{\bk'})
 \end{matrix}
 \right),
 \label{eq:Mexpli}
 \el
\eea
where $xx, yy$ elements correspond to upper or lower sign, respectively.
Furthermore the diagonal elements of  $M_{\la'\la}^{\alpha\alpha}$ matrices describe  intraband $(\la'=\la)$ and the nondiagonal ones 
interband $(\la'\neq\la)$ dipolar transitions between the Rashba bands.
The matrix elements for the nondiagonal cartesian indices are given in Appendix \ref{sec:Mnon}.
The complete dynamical response functions may now be obtained from Eq.~(\ref{eq:dynsus2})
using Eqs.~(\ref{eq:kernel-sus},\ref{eq:cohfac},\ref{eq:Mexpli}) and for the nondiagonal case 
Eq.(\ref{eq:nonexpli}) as input.\\

%
\begin{figure}
\includegraphics[width=\linewidth]{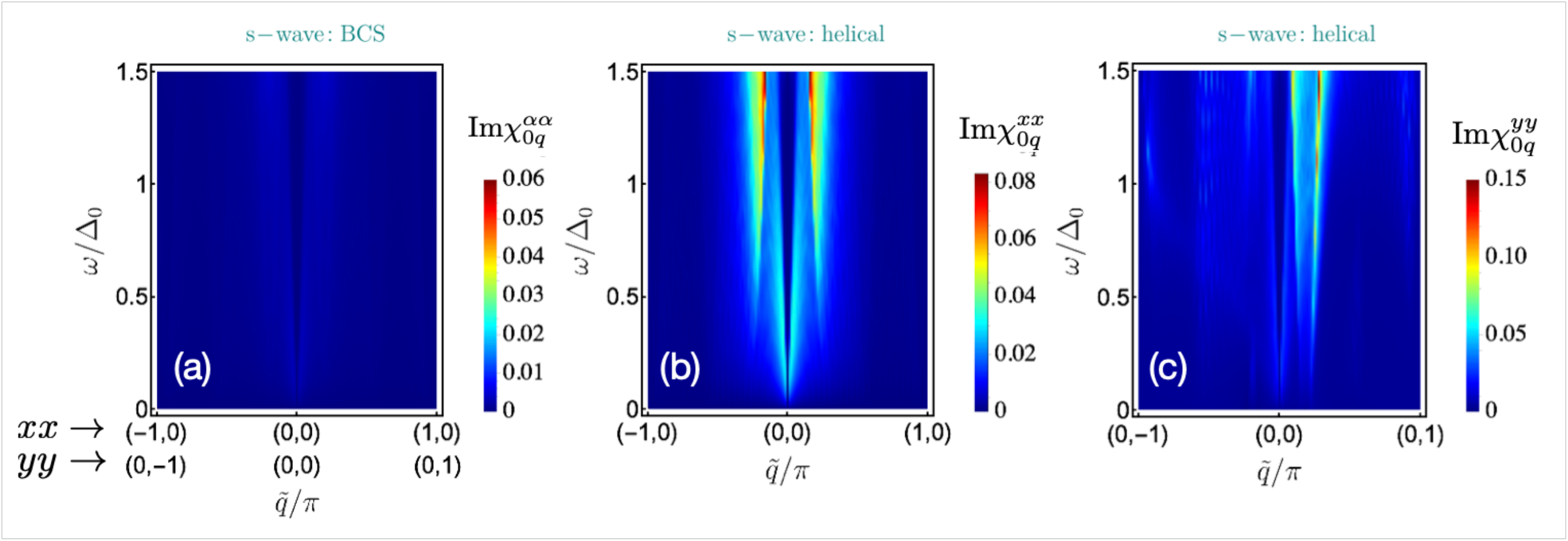}
\caption{Dispersive plot of magnetic excitations  using spin polarisation/momentum configurations $xx(10)$ and  $yy(01)$ of dynamical magnetic structure functions for s-wave case. In the BCS limit $(\bq=0)$ (a) the response for the two geometries is equal and vanishing due to $\omega<2\Delta_0$ (cf. Appendix \ref{sec:BCS-response}). In the helical phase (b,c) the low energy quasiparticle sheets (Fig.\ref{fig:BZcut}(i)) lead to non-vanishing dispersive excitations with strong momentum and spin-space anisotropy complementary to Figs.\ref{fig:BZcut},\ref{fig:linecut}.}
\label{fig:disp}
\end{figure}
%

The above generalized helical expressions reduce to the wellknown BCS results for the magnetic response \cite{bulut:96,norman:00,ismer:07,michal:11}
in the BCS limit $(b,q=0)$ which are given in Appendix \ref{sec:BCS-response} for comparison. If we restrict to the case
where $\bk,\bk'$ lie both in the segment with paired states (e.g. $E_{\bk\bq\la}>0, E_{\bk'\bq\la'}>0$)  then the terms in
 Eq.~(\ref{eq:kernel-sus})
may be consecutively interpreted as: quasiparticle scattering (first two terms), pair annihilation (third) and pair creation (fourth) terms.
For general $\bk,\bk'$~ one has to consider processes involving quasiparticles from the paired (p) as well as the unpaired (u) Fermi surface
segments. To simplify matters in this general case we consider the zero temperature limit when the Fermi function may be expressed by the step 
function according to  $f(E)=1-\Theta(E)=\Theta(-E)$. Then we obtain 
\be
\bl
\chi^{\al\bt}_{0\bq}(\tbq,\omega) =
\;
\;
&
\frac{1}{N}\sum_{\bk\la,\la'} M^{\al\bt}_{\la'\la}(\bk',\bk) 
\times
\\
\Bigl\{
&
\fs\tC^\bq_+(\bk\la\bk'\la')\bigl[
\frac{\Theta(E^+_{\bk\bq\la})-\Theta(E^+_{\bk'\bq\la'})}{\omega-(E^+_{\bk'\bq\la'}-E^+_{\bk\bq\la})+i\eta}-
\frac{\Theta(E^-_{\bk\bq\la})-\Theta(E^-_{\bk'\bq\la'})}{\omega+(E^-_{\bk'\bq\la'}-E^-_{\bk\bq\la})+i\eta}
\bigr]
+
\\
&
\fs\tC^\bq_-(\bk\la\bk'\la')\bigl[
\frac{\Theta(E^+_{\bk\bq\la})-\Theta(-E^-_{\bk'\bq\la'})}{\omega+(E^-_{\bk'\bq\la'}+E^+_{\bk\bq\la})+i\eta}+
\frac{\Theta(-E^+_{\bk'\bq\la'})-\Theta(E^-_{\bk\bq\la})}{\omega-(E^+_{\bk'\bq\la'}+E^-_{\bk\bq\la})+i\eta}
\bigr]
\Bigr\}
.
\label{eq:dynsus0}
\el
\ee
If we look at the numerators of the four terms in this equations we realize that the first two correspond
to quasiparticle scattering processes $\bk\leftrightarrow\bk'$ from paired (p) to unpaired (u)  FS segments and 
vice versa (p-u,u-p). whereas the third and fourth term are quasiparticle annihilation and creation respectively,
containing only processes between the paired (p-p) or unpaired (u-u) segments. 
The dynamical structure function for low temperature (without the Bose factor) which is proportional to 
the INS cross section \cite{jensen:91} is then obtained as
\be
S(\tbQ,\omega)=\sum_{\al\bt}(\delta_{\al\bt}-\hat{Q}_\al\hat{Q}_\bt)\frac{1}{\pi}
{\rm Im}
\chi^{\al\bt}_{0\bq}(\tbQ,\omega),
\label{eq:dynstruc}
\ee
where $\tbQ=\bk'-\bk=\tbq+\bK$ is the total momentum transfer with \bK~denoting a reciprocal lattice vector
and $\hat{\bQ}=\tbQ/|\tbQ|$ denoting the unit vector or direction of total momentum transfer. The prefactor projects out 
only scattering processes where the magnetic moment is perpendicular to $\tbQ$. By chosing various appropriate values
of the latter the individual susceptibility components, in particicualr the diagonal ones $(\al=\bt)$ on which we will focus 
in the discussion below can be accessed by INS.
\\

Finallly we may also consider the special case of the static staggered susceptibility components by setting 
$\omega=0$ in Eq.~(\ref{eq:kernel-sus}). After some rearrangements  we obtain $(\bar{\tau}=-\tau)$:
\be
\bl
\chi^{\al\bt}_{0\bq}(\tbq) =
\frac{1}{4N}\sum_{\bk\la\la'\tau} M^{\al\bt}_{\la'\la}(\bk',\bk) \Bigl\{
&
\tC^\bq_+(\bk\la\bk'\la')
\frac{\tanh\frac{\bt}{2}E^\tau_{\bk'\bq\la'}-\tanh\frac{\bt}{2}E^\tau_{\bk\bq\la}}{E^\tau_{\bk'\bq\la'}-E^\tau_{\bk\bq\la}}
+
\\
&
\tC^\bq_-(\bk\la\bk'\la')
\frac{\tanh\frac{\bt}{2}E^{\bar{\tau}}_{\bk'\bq\la'}+\tanh\frac{\bt}{2}E^\tau_{\bk\bq\la}}{E^{\bar{\tau}}_{\bk'\bq\la'}+E^\tau_{\bk\bq\la}}
\Bigr\}
.
\label{eq:statsus}
\el
\ee
We may further specify to $\tbq=0$ which is the homogeneous spin susceptibility. In this case the nondiagonal $(\la\neq\la')$ contributions are interband vanVleck terms with a large energy denominator whose modulus is $2|\al|k_F\gg 2\Delta_0$. Therefore they may be neglected compared to intraband terms $(\la'=\la)$. Then using
Using $\tC_+^\bq(\bk\la\bk\la)=1$ and $\tC_-^\bq(\bk\la\bk\la)=0$  we arrive at
\be
\chi^{\al\bt}_{0\bq}(0)\simeq\frac{1}{2N}\sum_{\bk\la\tau}M^{\al\bt}(\bk\la)
\Bigl(-\frac{\partial f}{\partial E^\tau_{\bk\bq\la}}\Bigr)
=\frac{\beta}{4}\frac{1}{2N}\sum_{\bk\la\tau}M^{\al\bt}(\bk\la)\Bigl(\frac{1}{\cosh^2\frac{\beta}{2}E^\tau_{\bk\bq\la}} \Bigr)
.
\ee
Where the diagonal helical matrix elements now simplify to 
\bea
M^{xx}(\bk\la)=\fs(1-\cos\theta_\bk);
\quad
M^{yy}(\bk\la)=\fs(1+\cos\theta_\bk);
\quad
M^{zz}(\bk\la)=0
.
\label{eq:matsimpl}
\eea
In the parabolic band approximation (for $\mu\ll D_c$) with a 2D DOS $N^0_\la=m^*k^\la_F/2\pi$ and effective mass $m^*=2/D_c$ and Fermi vector $k_F^\la=k_F-\la k_0$, $k^0_F=(2m^*\mu)^\fs$ the homogeneous susceptibility may approximately be written as
\bea
\chi^{\al\al}_{0\bq}(0,T)=\sum_{\la\tau}N^0_\la\int \frac{d \theta_\bk}{2\pi} M^{\al\al}(\theta_\bk\la) Y_{q\la}^\tau(\theta_\bk,T);\;\;\; Y^\tau_{\bq\la}(\theta_\bk,T)=
\frac{1}{4\pi}\int\frac{d\vare_\la}{\cosh^2\frac{\beta}{2}E^\tau_{\bk\bq\la}}
,
\eea
where $Y^\si_{q\la}(T,\theta_\bk)$ is a  generalized angular resolved Yosida function \cite{mineev:99}. The static homogeneous susceptibility $\chi^{\al\al}_{0\bq}(\tbq)$ describes the temperature dependence of the NMR Knight shift \cite{vorontsov:06,thalmeier:22} of the singlet superconductor (for $\alpha=0$)  in the helical phase. Because of Eq.~(\ref{eq:matsimpl}) $xx$ and $yy$ components are equivalent and the $zz$ component vanishes.
For plotting the temperature dependence of the intra-band contribution we use a phenomenological temperature dependence of the helical gap $\Delta_q$ given by $\Delta_q(t)=\Delta_q\tanh[1.74\sqrt{\frac{1-t}{t}}]$ where $t=T/T_c(H)$ is the reduced temperature. The comparison of $\chi^{\al\al}_{0\bq}(0,T)$ in the BCS $(\bq=0)$ and helical  $(\bq\neq 0)$ case in the interval $t\in [0,1]$ is shown in Fig.~\ref{fig:statsus}.

%
\begin{figure}
\begin{center}
\includegraphics[width=0.65\linewidth]{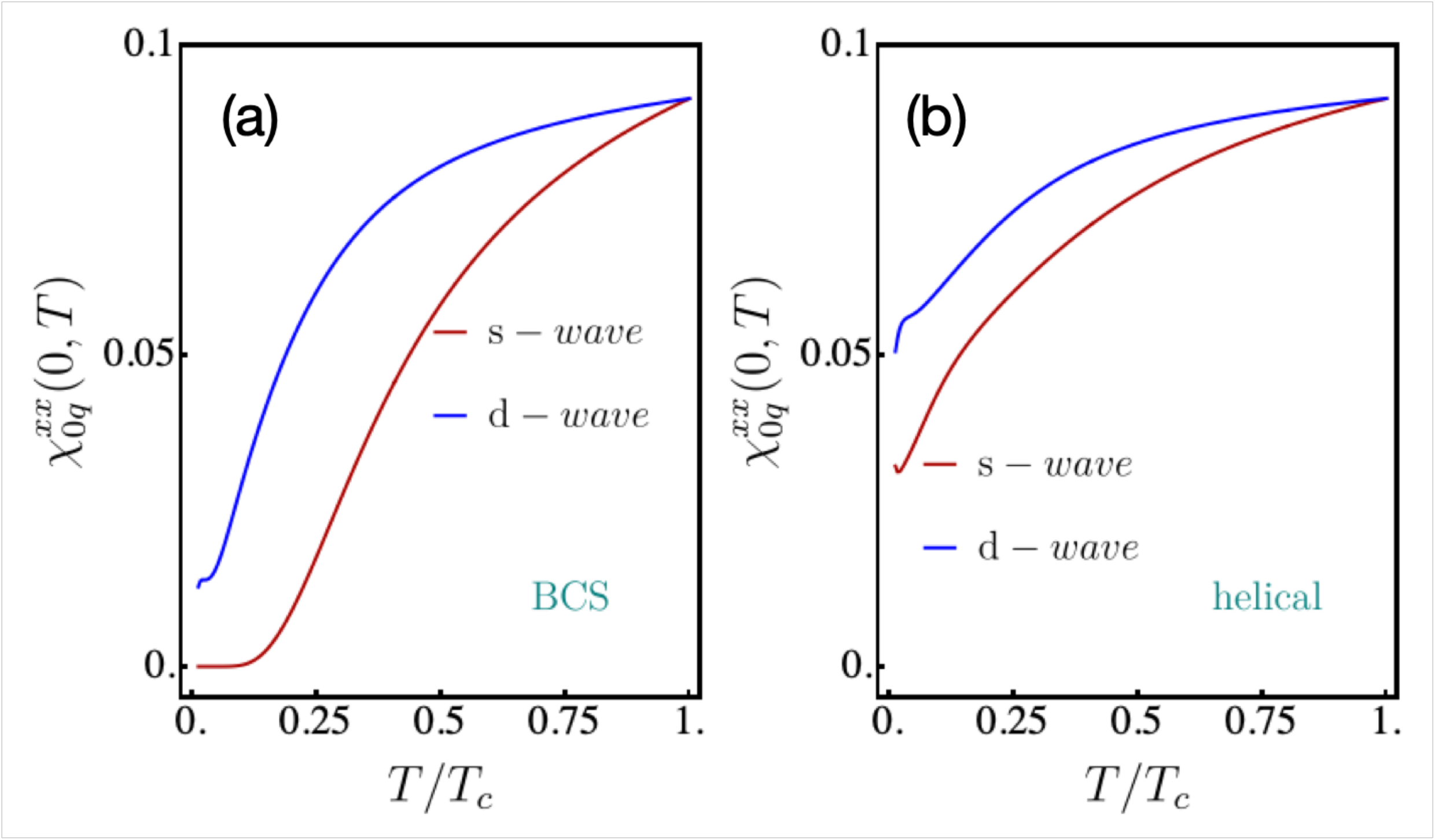}
\caption{Temperature dependence of static homogeneous susceptibility contributing to NMR Knight shift for BCS and helical phase.
The large nonzero value for $T\rightarrow 0$ in (b) is due to normal quasiparticles in depaired momentum space region.}
\label{fig:statsus}
\end{center}
\end{figure}
%

\section{Discussion of numerical results: the magnetic spectral functions}
\label{sect:discussion}

Here we discuss typical numerical results for the magnetic spectrum that may be obtained from the theory developed in the 
previous sections. In particular we focus on the resulting various momentum-space and cartesian spin space anisotropies of the response function with respect to the direction of the field  $\bb=b\hat{{\bf x}}$ and the corresponding overall pair momentum direction $\bq=q\hat{{\bf y}}$. In this discussion we have to restrict to results for the small field range $b\ll\alpha$ where the assumption of field-independent 
spin texture of Rashba states is still acceptable. This is the basis for the simplified gap models \cite{kaur:05} employed here.
For the feasibility of numerical computations we had to use a sizable gap amplitude of $\Delta_0/t=0.2$. We use the s- and d- wave gap models. We stress that this designation refers to the limiting case of $\alpha=0$ as mentioned in Sec.~\ref{subsect:supermodel}.
In the majority of results presented here we use a Rashba parameter value $\alpha=0.6t$ and field strength $b=0.9\Delta_0$ (i.e. $b/\alpha =0.3$). Furthermore the chemical potential is set to $\mu=\mu_{TB}= -2.8t$ (or $\mu_P=1.2t$) to achieve quasi- circular Rashba Fermi surfaces. For  $b=0.9\Delta_0$, we found  ($q/\pi=0.011$;  $\Delta_q=0.75  \Delta_0$) and ($q/\pi=0.015$; $\Delta_q=0.85 \Delta_0$) for s-wave and d-wave, respectively.
\\  

In Fig.~\ref{fig:spectral} we show the field dependence of $q,\Delta_q$ in the small field region (a,b) and the associated change in the 
quasiparticle sheets represented by the spectral function for both gap symmetries. In the BCS case the fully gapped s-wave state
has none while in  the d-wave case the sheets for both Rashba bands (inner and outer circle)  appear first around the nodal directions.
In the helical phase with finite overall pair momentum $q$ large quasiparticle sheets exist around the depaired momentum space regions perpendicular to  the field direction. It is obvious that the fourfold symmetry of the BCS state is broken by the distinguished \bq -direction and this should be visible in the dynamical magnetic structure function as a momentum-space anisotropy of the response. Because in the Rashba state with dominating $\alpha$ spin and momentum directions are locked this should also be transfered to a spin-space anisotropy, expressed by non-equivalence of $\chi_0^{\alpha\alpha}(\bq,\omega)$ for the cartesian directions $\alpha=x,y,z$. While the former effect is already observed in the centrosymmetric FF state \cite{thalmeier:22}, the latter is characteristic for the helical Rashba SC due to the spin-momentum locking.\\

This is nicely illustrated by the constant-$\omega$  $(=0.12t = 0.6\Delta_0)$ cut of the spectrum (imaginary part of the dynamical susceptibility) in the panels of Fig~\ref{fig:BZcut}. It shows the BCS and helical d-wave case in the top and center row and the helical s-wave case in the bottom row (in-the BCS s-wave case there is no intensity at finite $\omega<2\Delta_0$ due to the full gapping (Appendix \ref{sec:BCS-response})). The spectral functions are depicted on the left column and the $xx,yy,zz$ dynamical magnetic response in the three consecutive columns. We notice that in the BCS case the combined spin/momentum space rotational symmetries
\be
\bl
\chi_{00}^{xx}(R_\frac{\pi}{2}\tbq.\omega)=\chi_{00}^{yy}(\tbq,\omega);
\quad
\chi_{00}^{yy}(R_\frac{\pi}{2}\tbq,\omega)=\chi_{00}^{xx}(\tbq,\omega);
\quad
\chi_{00}^{zz}(R_\frac{\pi}{2}\tbq,\omega)=\chi_{00}^{zz}(\tbq,\omega),
\label{eq:BZsymmetry}
\el
\ee
hold for any $\tbq$ in the BZ where $R_\frac{\pi}{2}$ denotes the rotation by $\frac{\pi}{2}$ around $\tilde{q}_z$- axis. These symmetries are all violated in the helical phases where the common pair momentum  $\bq=q\hat{{\bf y}}$ leads to the anisotropic response 
depicted in (f-h) and (j-l). This spin/momentum space anisotropy in the $xy$ and $q_xq_y$ planes is a fingerprint of the helical SC phase. It  should be accessible experimentally by constant-$\omega$ scans of the INS intensity in the BZ. The various cartesian spin channels for $\alpha=x,y,z$ may be selected by using a suitable total momentum transfer according to Eq.~(\ref{eq:dynstruc}).\\

There are complementary ways of presenting these anisotropies characteristically appearing in the helical phase. One possibility is to make cuts through the BZ and plot the dynamical response along that direction. This is shown in Fig.~\ref{fig:linecut} for $(10)$ and $(01)$ directions and $xx,yy$ comonents in the s-wave case. For BCS limit we have again pairwise equivalence of response according to Eq.(\ref{eq:BZsymmetry}) as evident from (a,d). In the helical phase these symmetries are destroyed as is shown by the further cuts (b,c) and (e,f).
Another complementary presentation of the helical anisotropies is shown in the dispersive plots of Fig.~\ref{fig:disp}, again for the s-wave case. In the BCS limit of (a) the dynamical response in $xx(10)$ $yy(01)$ is equivalent  and vanishing for $\omega<2\Delta_0$. For the helical case in the same configurations the dispersive magnetic excitations are clearly present due to low energy quasiparticles in the depaired momentum space sectors but strongly anisotropic for the two cases, in agreement with previous presentations.\\

Another interesting result of this investigation is the temperature dependence of the static spin susceptibility which in principle determines the NMR Knight shift. It is shown in Fig.~\ref{fig:statsus} in comparison for BCS and helical states. Note that according to Sec.\ref{sec:helical-response} $xx$ and $yy$ susceptibility are equivalent for $\tbq=0$ for both BCS {\it and} helical state, therefore we plot only the $xx$ component. In the BCS case the well known exponential and power law behaviour of the susceptibility for the s- and d-wave cases are observed. The appearance of large sheets of  low energy quasiparticles in the helical phase leads to large residual low temperature susceptibility which should be observable. We note that this happens although the helical order parameter does {\it not} have nodes in real space as it is true, e.g. for the LO phase in the case without inversion symmetry breaking \cite{matsuda:07}.

\section{Summary and Conclusion}
\label{sec:conclusion}

This work completes our previous investigations of spectroscopic (QPI, INS) properties of superconducting states with finite momentum pairing 
\cite{akbari:16,akbari:22,thalmeier:22}. We have derived the general expressions of dynamical magnetic response in noncentrosymmetric superconductors with Rashba spin orbit coupling that may stabilize the helical phase in an applied field. We have shown which type of quasiparticle excitations appear in the response functions and derived the corresponding anomalous coherence factors and matrix elements due to helical spin textures of the Rashba bands. The obtained expressions of the response function generalize the known results for the centrosymmetric BCS case to the two-band Rashba case with spin-momentum locking under the presence of applied fields which are small compared to the Rashba coupling energy.

As a major result we demonstrated that the combined spin-/momentum space symmetries of the BCS phase response is broken in the helical phase. It leads to characteristic spin/momentum asymmetries and anisotropies which can in principle be detected by inelastic neutron scattering experiments. In particular the constant-$\omega$ cuts of the magnetic spectrum by scanning through the BZ should give a fingerprint of the helical phase. From the type of observed anisotropies it should be possible to conclude about the direction of the \bq- pair momentum vector and observe its dependence on applied field direction. A further possibility is to look for the frequency dependence at certain $\tbq$ momentum transfer and whether spin resonance excitations may form as a consequence of quasiparticle interactions similar as has been predicted for the centrosymmetric case \cite{thalmeier:22}.

Finally we investigated the generic temperature dependence of the static spin susceptibillity important to know for NMR experiments.
We demonstrated that in the helical phase a large residual low temperature spin susceptibility remains which signifies the appearance
of unpaired states in this phase, although these are not due to real space nodal planes of the order parameter itself.
In summary our investigation gives a solid theoretical foundation for magnetic spectroscopy of the helical Rashba superconducting state.



\appendix

\section{Spin operator matrix elements for the nondiagonal cartesian cases}
\label{sec:Mnon}
 
 In a similar manner as in Sec.~\ref{sec:helical-response} the matrix elements $M^{\al\bt}_{\la\la'}$ for $\al\neq\bt$ may be derived as
\be
\bl
M^{xy}(\bk'\bk)
=&
\fs \left(
\begin{matrix}
-\sin(\theta_\bk+\theta_{\bk'})& \sin(\theta_\bk+\theta_{\bk'}) \cr
 \sin(\theta_\bk+\theta_{\bk'}) & -\sin(\theta_\bk+\theta_{\bk'})
 \end{matrix}
 \right),
 \\
 M^{xz}(\bk'\bk)  
=&
\fs\left(
\begin{matrix}
i(\cos\theta_\bk-\cos\theta_{\bk'})& -i(\cos\theta_\bk+\cos\theta_{\bk'}) \cr
i(\cos\theta_\bk+\cos\theta_{\bk'}) & -i(\cos\theta_\bk-\cos\theta_{\bk'})
 \end{matrix}
 \right),
 \\
 M^{yz}(\bk'\bk)
 =&
\fs\left(
\begin{matrix}
i(\sin\theta_\bk-\sin\theta_{\bk'})& -i(\sin\theta_\bk+\sin\theta_{\bk'}) \cr
i(\sin\theta_\bk+\sin\theta_{\bk'}) & -i(\sin\theta_\bk-\sin\theta_{\bk'})
\end{matrix}
\right).
\label{eq:nonexpli}
\el
\ee
 They appear in the expressions for the nondiagonal susceptibility components $\chi_0^{\al\bt}(\tbq,\omega)$.
 
 \section{Magnetic response function in the BCS limit}
 \label{sec:BCS-response}

 The diagonal cartesian susceptibilities $(\alpha=\beta)$ of  Eqs.~(\ref{eq:dynsus2},\ref{eq:kernel-sus}) may be rewritten by using the 
 symmetry of matrix elements $M^{\alpha\alpha}_{\la'\la}(\bk',\bk)$ and coherence factors $\tC^\bq_\pm(\bk\la\bk'\la')$ against interchange of primed and unprimed arguments. Furthermore, since the summation over both is identical, the first two terms  of  Eqs.~(\ref{eq:dynsus2},\ref{eq:kernel-sus}) may be contracted into one term. Then in the BCS case $(b,q=0)$ when quasiparticle bands are given by $E^\tau_{\bk\bq\la}=E_{\bk\la}=(\vare_{\bk\la}^{02}+\De_\bk^2)^\fs$ with $\vare_{\bk\la}^0$ defined by Eq.~(\ref{eq:Rdisp0}) we obtain the simplified result
 \be
 \bl
 \chi^{\al\al}_{0}(\tbq,\omega) =
 &
 \frac{1}{N}\sum_{\la\la'}
\sum_{\bk}
M^{\al\al}_{\la'\la}(\bk',\bk)
\Bigl\{\tC_+(\bk\la,\bk'\la')
\frac{f(E_{\bk'\la'})-f(E_{\bk\la})}{\omega-(E_{\bk'\la'}-E_{\bk\la})+i\eta}
\\
&
+\fs\tC_-(\bk\la\bk'\la')\bigl[
\frac{1-f(E_{\bk'\la'})-f(E_{\bk\la})}{\omega+(E_{\bk'\la'}+E_{\bk\la})+i\eta}+
\frac{f(E_{\bk'\la'})+f(E_{\bk\la})-1}{\omega-(E_{\bk'\la'}+E_{\bk\la})+i\eta}
\bigr]\Bigr\}
.
 \el
 \ee
This  is a generalisation of BCS expressions given in Refs.~\cite{bulut:96,norman:00,ismer:07,michal:11} to the magnetic response for 
the two-band Rashba-BCS superconductor. The anomalous coherence factors of Eq.~(\ref{eq:cohfac}) now simplify to
\be
\tC_\pm(\bk\la\bk'\la')=\fs\bigl[
1\pm \frac{\vare^0_{\bk\la}\vare^0_{\bk'\la'}+\Delta_{\la}^\bk\Delta_{\la'}^{\bk'}}
{E_{\bk\la}E_{\bk'\la'}}\bigr].
\label{eq:cohfacBCS}
\ee
 In the low temperature limit and for positive frequencies
 only the last term survives leading to
 \be
 \bl
 \chi^{\al\al}_{0}(\tbq,\omega) =-\frac{1}{N}\sum_{\la\la'}
\sum_{\bk}
M^{\al\al}_{\la'\la}(\bk',\bk)
 \frac{\fs\tC_-(\bk\la\bk'\la')}{\omega-(E_{\bk'\la'}+E_{\bk\la})+i\eta}
 .
 \el
 \ee
 The spectrum of the $T=0$ diagonal BCS response functions is then given by ($\tbq\in 1^{st}$ BZ)
 \be
 \bl
 \hat{S}_{\al\al}(\tbq,\omega)=\frac{1}{\pi}
 {\rm Im}
 \chi^{\al\al}_{0}(\tbq,\omega)
 =\frac{1}{N}\sum_{\la\la'}
\sum_{\bk}
M^{\al\al}_{\la'\la}(\bk',\bk)
\fs\tC_-(\bk\la\bk'\la')\delta[\omega-(E_{\bk'\la'}+E_{\bk\la})]
,
 \el
 \ee
 which contains both intra- and interband transitions between the Rashba-split quasiparticle bands. For $\tbq=0$ i.e. $\bk=\bk'$ we obtain
 \be
 \bl
 \hat{S}_{\al\al}(\omega)
 =\frac{1}{N}\sum_{\la\la'}
\sum_{\bk}
M^{\al\al}_{\la'\la}(\bk,\bk)
\fs\tC_-(\bk\la\bk\la')\delta[\omega-(E_{\bk\la'}+E_{\bk\la})]
.
 \el
 \ee
 The threshold values for this spectrum are different for intra- and interband transitions. For the former $(\la'=\la)$
 it is given by $2\Delta_0$ for the latter  $(\la'\neq\la)$ by $\omega^\pm_0=|\alpha|k^\pm_F\gg 2\Delta_0$.
 Therefore the threshold of the total spectrum is the intraband threshold $2\Delta_0$.


\bibliography{ReferencesRH}

\end{document}